\documentclass[prd,twocolumn,showpacs,nofootinbib, aps,10pt,superscriptaddress,amsmath,amssymb]{revtex4-1}
\usepackage{amssymb}
\usepackage{epsfig}
\usepackage{graphicx}
\usepackage{subfigure}
\usepackage{dcolumn}
\usepackage{bm}
\usepackage[usenames ,dvipsnames]{xcolor}
\usepackage{slashed}

\begin{document}

\title{Comment on ``Hearing the signal of dark sectors with gravitational wave detectors'' \\
{[Phys.\ Rev.\ D {\bf 94}, no. 10, 103519 (2016)]}}

\author{Da~Huang}\email{dahuang@fuw.edu.pl} 
 \affiliation{Institute of Theoretical Physics, Faculty of Physics, University of Warsaw, Pasteura 5, 02-093 Warsaw, Poland}
\author{Bo-Qiang Lu}\email{bqlu@itp.ac.cn}
\affiliation{State Key Laboratory of Theoretical Physics, \\
Institute of Theoretical Physics, Chinese Academy of Sciences, Beijing, 100190, P.R. China
}

\date{\today}
\begin{abstract}
We revisit the calculation of the gravitational wave spectra generated in a classically scale-invariant $SU(2)$ gauge sector with a scalar field in the adjoint representation, as discussed by J.~Jaeckel, {\it et~al}. The finite-temperature potential at 1-loop level can induce a strong first-order phase transition, during which gravitational waves can be generated. With the accurate numerical computation of the on-shell Euclidean actions of the nucleation bubbles, we find that the triangle approximation employed by J.~Jaeckel, {\it et~al.} strongly distorts the actual potential near its maximum and thus greatly underestimates the action values. As a result, the gravitational wave spectra predicted by J.~Jaeckel, {\it et~al.} deviate significantly from the exact ones in peak frequencies and shapes.
\end{abstract}

\maketitle

The discovery of gravitational waves (GWs) by advanced LIGO (aLIGO)~\cite{Abbott:2016blz} is a great step to confirm the Einstein's General Relativity, and it is more exciting to exploit the GWs as a probe to the dark sector beyond the Standard Model (SM) of particle physics. Following this line of thinking, Joerg Jaeckel, {\it et al.} in Ref.~\cite{Jaeckel:2016jlh} studied the GW production in a classically scale-invariant sector in which a strong first-order phase transition (PT) occurs via the one-loop potential at finite temperatures, finding that the produced GWs can be detected by the near-future experiments including fifth phase of aLIGO. In their calculation of the GW spectra, J.~Jaeckel {\it et al.} employed the triangle approximation to the one-loop finite-temperature scalar potential in order to obtain the on-shell Euclidean action analytically. However, as will be shown in the present work, this triangle approximation is not good enough for this classically scale-invariant model, since it significantly distorts the potential. As a result, the GW spectra given in Ref.~\cite{Jaeckel:2016jlh} deviate considerably from the exact ones in the main sources, peak frequencies and shapes.

The model studied in Ref.~\cite{Jaeckel:2016jlh} is quite simple. It is only composed of a classically scale-invariant $SU(2)$ gauge sector with one scalar field in the adjoint representation\footnote{J.~Jaeckel {\it et al.} in Ref.~\cite{Jaeckel:2016jlh} also investigated a classically scale-invariant $U(1)$ gauge group with a scalar of unit charge. But their numerical results and plots were based only on the $SU(2)$ gauge theory presented in the main context.}. Following the assumptions and notations in Ref.~\cite{Jaeckel:2016jlh}, the zero-temperature one-loop potential is given by 
\begin{eqnarray}\label{Vzero}
V(\phi) = \frac{n_W}{64\pi^2} g_D^4 \phi^4 \left[-\frac{1}{2} + \log\left( \frac{\phi^2}{w^2} \right) \right]\,,
\end{eqnarray}
where $\phi$ is the real Higgs boson in the original $SU(2)$ triplet left after the spontaneous breaking of $SU(2)$ gauge symmetry, and $w$ is the vacuum expectation value (VEV) of the scalar field $\phi$ at the true potential minimum. In Eq.~(\ref{Vzero}), We have only taken into account the leading-order gauge boson contribution to the scalar potential with $g_D$ as the gauge coupling constant and $n_W = 6$ as the degrees of freedom (dofs) for massive gauge bosons. At a temperature $T$, there is an additional one-loop finite-temperature correction given by
\begin{eqnarray}
\Delta V_T &=&  \frac{n_W T^4}{2\pi^2}\int^\infty_0 dq q^2 \times \nonumber\\
&& \log\left( 1 - \exp\left[-\sqrt{q^2 + m_W^2(\phi)/T^2}\right] \right)\,,
\end{eqnarray}
where $m_W(\phi) = g_D \phi$ is the $\phi$-dependent mass of massive vector bosons. The total effective scalar potential at finite temperatures in this model is given by
\begin{eqnarray}\label{Vtot}
V_T (\phi) = V(\phi) + \Delta V_T(\phi)\,,
\end{eqnarray}
which is the starting point for our discussion of the first-order PT and the GW production.

A typical first-order PT goes via the bubble nucleation, in which the rate is defined as~\cite{Espinosa:2008kw,Caprini:2015zlo}
\begin{eqnarray}
\Gamma(t) = A(T) e^{-S_3(\phi_{\rm cl})/T}\approx \kappa_3 T^4 e^{-S_3/T}\,,
\end{eqnarray}
where $S_3$ denotes the three-dimensional (3d) on-shell Euclidean action of a bubble solution, and $\kappa_3 = [S_3/(2\pi T)]^{3/2}$. Following Ref.~\cite{Espinosa:2008kw}, we define the average bubble nucleation number per Hubble volume until a temperature $T$ as follows:
\begin{eqnarray}
P(T) = \int^{T_c}_T \kappa_3 \frac{d\tilde{T}}{\tilde{T}}\frac{\tilde{T}^4}{H^4} e^{-S_3(\tilde{T})/\tilde{T}} \approx \frac{T^4}{H^4}e^{-S_3(T)/T}\,,
\end{eqnarray}
where $H$ denotes the Hubble parameter determined by the Friedmann equation
\begin{eqnarray}\label{Friedmann}
H^2 = \frac{8 G \pi^2 \rho}{3} \simeq \frac{8\pi^3 g_* T^4}{90 M^2_{\rm Pl}}\,,
\end{eqnarray}
in which $M_{\rm Pl} = 1.22\times 10^{19}$~GeV is the Planck mass, and we have used the relation $\rho \approx \rho_{\rm rad} = g_* \pi^2 T^4/30$ in the radiation domination epoch. Hence, the first-order PT occurs at the nucleation temperature $T_n$ when the probability to nucleate a bubble per Hubble volume reaches 1, $P(T_n)\gtrsim 1$. With the Friedmann equation in Eq.~(\ref{Friedmann}), this first-order PT criterion can be transformed into~\cite{Espinosa:2008kw}
\begin{eqnarray}\label{critPT}
\frac{S_3(T_n)}{T_n} \lesssim 4 \ln\left(\frac{T_n}{H}\right) = 2\ln\frac{90 M_{Pl}^2}{8\pi^3 g_* T_n^2}  \,,
\end{eqnarray}
where the effective number of relativistic dofs is taken as the SM value $g_* = 106.75$. In this work, we focus on the dark PT with the nucleation temperature around ${\cal O}({\rm TeV} \sim {\rm PeV})$, so that the criterion for a first-order PT can be estimated to be in the range $110 \lesssim S_3/T_n \lesssim 146$. For the later convenience, $S_3/T_n = 140$ is chosen as our criterion to determine the nucleation temperature.

GWs can be produced along with a violent first-order PT, so that the typical temperature for GW generation $T_*$ is usually taken to be $T_* \simeq T_n$. The stochastic GWs produced in this way are usually characterized by the following two dimensionless quantities~\cite{Caprini:2015zlo}:
\begin{eqnarray}\label{ab}
\frac{\beta}{H_*} = T \frac{d}{dT}\left(\frac{S_3}{T}\right)\Big|_{T=T_*}\,, \quad \quad \alpha = \frac{\rho_{\rm vac}}{\rho_{\rm rad}(T_*)}\,,
\end{eqnarray}
where $\beta$ is roughly the inverse time duration of the PT, $\rho_{\rm vac}$ denotes the released vacuum energy density, and $H_*$($\rho_{\rm rad}(T_*)$) is the Hubble parameter (thermal bath energy density) at $T_*$. However, when there is a significant supercooling during the PT, the formulae above are not valid and GWs are mostly radiated at a temperature after reheating, $T_* \approx T_{\rm reh}\gg T_n$. The formulae for $\alpha$ and $\beta/H_*$ are modified as follows~\cite{Caprini:2015zlo}
\begin{eqnarray}\label{abSuper}
\frac{\beta}{H_*} = \frac{H(T_n)}{H_*} T_n \frac{d}{dT}\left(\frac{S_3}{T}\right)\Big|_{T=T_n}\,\quad \alpha = \frac{\rho_{\rm vac}}{\rho_{\rm rad}(T_{n})}\,.
\end{eqnarray}
Now we assume that the reheating is so fast that the released vacuum energy can be transferred into the thermal bath energy without any loss, which leads to $H(T_n) = H_*$ even if $T_n\ll T_*$. In this case, the definitions of $\alpha$ and $\beta/H_*$ in Eq.~(\ref{abSuper}) are reduced to the normal ones in Eq.~(\ref{ab}). Thus, we do not distinguish these two definitions any more in the following.

For the scalar potential in Eq.~(\ref{Vtot}), the 3d Euclidean action is given by
\begin{eqnarray}\label{actionCW}
S_3 &=& \int d^3 {\bf x} \left[\frac{1}{2}\partial_i \phi({\bf x}) \partial_i \phi({\bf x}) + V_T(\phi) \right]\nonumber\\
 &=& 4\pi \int^\infty_0 r^2 dr\left[\frac{1}{2}\dot{\phi} +V_T(\phi)\right]
\end{eqnarray}
where ${\bf x}$ denotes the 3d Euclidean coordinate with the subscript $i$ as its index. Since we pursue spherically symmetric bubble solutions, the on-shell action can be reduced to the integral over the radial coordinate $r$ as in the second equality. By minimizing the action, we can obtain the following differential equation
\begin{eqnarray}\label{DiffEq}
\frac{d^2 \phi}{d r^2} + \frac{2}{r}\frac{d\phi}{dr} - \frac{dV_T}{d\phi}=0\,,
\end{eqnarray}
with the boundary conditions
\begin{eqnarray}\label{Bound}
\frac{d\phi}{dr}\Big|_{r = 0} = 0\,, \quad \quad \phi(\infty) = 0\,.
\end{eqnarray}
This equation can be solved by utilizing the overshooting-undershooting method, and the on-shell Euclidean action $S_3$ can be calculated by integrating over the obtained bubble profile. In this work, we employ the code \texttt{CosmoTransitions}~\cite{Wainwright:2011kj} to accurately perform such computations.

\begin{figure}[ht]
\includegraphics[scale = 0.42]{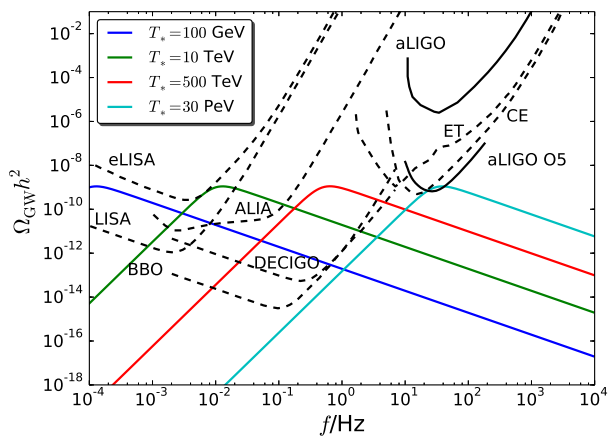}
\caption{The blue, green, red and cyan solid curves from left to right represent the gravitational wave spectra predicted with $g_X = 0.6$ and $T_* = 100$~GeV, 10~TeV, 500~TeV and 30~PeV, respectively. We also show the sensitivities of aLIGO as well as its fifth phase aLIGO-O5 as black solid curves, and the proposed experiments LISA, eLISA, BBO, ALIA, DECIGO, ET and CE as black dashed curves. }\label{GWCW}
\end{figure}

In order to determine the nucleation temperature $T_n$, we follow Ref.~\cite{Jaeckel:2016jlh} to define the dimensionless potential:
\begin{eqnarray}\label{VhatCW}
\hat{V}(\gamma, \Theta) &\equiv& \frac{V_T(\phi)}{g_X^4 v^4} \nonumber\\
 &=& \frac{6}{64\pi^2} \gamma^4 \left(\log\gamma^2 -\frac{1}{2}\right)+ \frac{6\Theta^4}{2\pi^2} \times \nonumber\\
 && \int^\infty_0 dq q^2 \log\left( 1 - \exp\left(-\sqrt{q^2 +\gamma^2/\Theta^2}\right) \right) \,,\nonumber\\
\end{eqnarray}
where we have defined the rescaled field $\gamma = \phi/w$ and rescaled temperature $\Theta = T/(g_D w)$. We can also define the associated dimensionless 3d action
\begin{eqnarray}\label{actionCW1}
\tilde{S}_3 = \int d^3 {\bf z} \left[\frac{1}{2}\partial_i \gamma({\bf z}) \partial_i \gamma({\bf z}) + \hat{V}_T(\gamma) \right]\,.
\end{eqnarray}
It is easy to show that under the coordinate transformation ${\bf x} = {\bf z}/(g_D w)$, the dimensionless quantities $S_3/T$, $\beta/H_*$, and $\alpha$ can be expressed in terms of the above rescaled quantities:
\begin{eqnarray}
&& \frac{S_3}{T} = \frac{1}{g_D^3} \frac{\tilde{S}_3}{\Theta}\,, \quad \frac{\beta}{H_*} = \frac{1}{g_D^3} \Theta \frac{d}{d\Theta} \left(\frac{\tilde{S}_3}{\Theta}\right)\bigg|_{\Theta=\Theta_*}\,,\nonumber\\
&& \alpha = \frac{30}{\pi^2 g_* \Theta^4} \left[ \hat{V}(0) - \hat{V}_{\rm min} \right]\bigg|_{\Theta = \Theta_*}\,,
\end{eqnarray}
where $\hat{V}_{\rm min}$ ($\hat{V} (0)$) denotes the dimensionless vacuum energy density at the true (false) vacuum. It is interesting to note that all of these quantities do not depend the overall scale $w$ in the theory. Hence, for $g_D = 0.6$ as in Ref.~\cite{Jaeckel:2016jlh}, the dimensionless nucleation temperature can be determined to be $\Theta_n \approx 0.01$ for $S_3/T_n = \tilde{S}_3/(g_D^3 \Theta_n) = 140$. Other quantities related to the GW production can also be calculated to be $\beta/H_* = 33.68$ and $\alpha=1.41\times 10^4 \gg 1$, which indicates that there is a huge supercooling in this dark first-order PT. The situation is very similar to the nearly conformal models studied in the context of holographic PT in Ref.~\cite{Konstandin:2011dr,Konstandin:2010cd}. In order to determine the GW spectra, we need further to specify its main production processes. Due to the significant supercooling, it is known~\cite{Caprini:2015zlo} that the PT should proceed via runaway bubbles in the vacuum, which predicts that most of the latent heat is transferred into the kinetic energy of the scalar field. The bubble wall velocity approaches the speed of light $v_b = 1$, and the dominant contribution to GWs comes from bubble collisions, rather than from sound waves and the magnetohydrodynamic (MHD) turbulence as argued in Ref.~\cite{Jaeckel:2016jlh}. Thus, the efficiency parameters for the scalar field and the plasma are given by $\kappa_\phi \approx 1$ and $\kappa_v \to 0$. Furthermore, the prediction of GW spectra requires the overall scale of the PT. Following Ref.~\cite{Jaeckel:2016jlh}, it is more convenient to use the GW production temperature $T_*$ to represent this scale, since it is $T_*$ that appears in the GW spectra formulae for three mechanisms given in Ref.~\cite{Caprini:2015zlo}. As mentioned before, in the case with a large supercooling, $T_*$ should be taken as the plasma temperature after the reheating, $T_*= T_{\rm reh}$. In Fig.~\ref{GWCW}, we show the predicted GW spectra for $g_X = 0.6$ and several typical GW production temperatures $T_* = 100$~GeV, 10~TeV, 500~TeV and 30 PeV, as well as the sensitivities of existent aLIGO with its fifth phase~\cite{TheLIGOScientific:2016wyq} and several proposed detectors, all of which are obtained from Ref.~\cite{Moore:2014lga}.

It is clear that the GW spectra given in Fig.~\ref{GWCW} are quite different from those in Fig.~6 and 7 of Ref.~\cite{Jaeckel:2016jlh}. Firstly, our predicted GW peak frequency is much lower than that in Ref.~\cite{Jaeckel:2016jlh} for a fixed production temperature $T_*$, which is mainly due to the different values of $\beta/H_*$. In Ref.~\cite{Jaeckel:2016jlh}, $\beta/H_*$ is chosen to be around $150$, while in our case its value is $33.68$. Secondly, the shapes of the GW spectra are completely dissimilar, which reflects the differences in dominant GW production mechanisms. In our case, vacuum bubble collisions contribute dominantly, $\Omega_{\rm GW} h^2 = \Omega_{\phi} h^2$, while the calculations by Jaeckel, {\it et al}. favored the sound waves and MHD turbulences as main GW sources. In the following, we would like to argue that the above distinctions in the GW predictions are originated from the problematic application of the triangle approximation to the scalar potentials and the on-shell bubble actions in Ref.~\cite{Jaeckel:2016jlh}. 

\begin{figure}[ht]
\includegraphics[scale = 0.38]{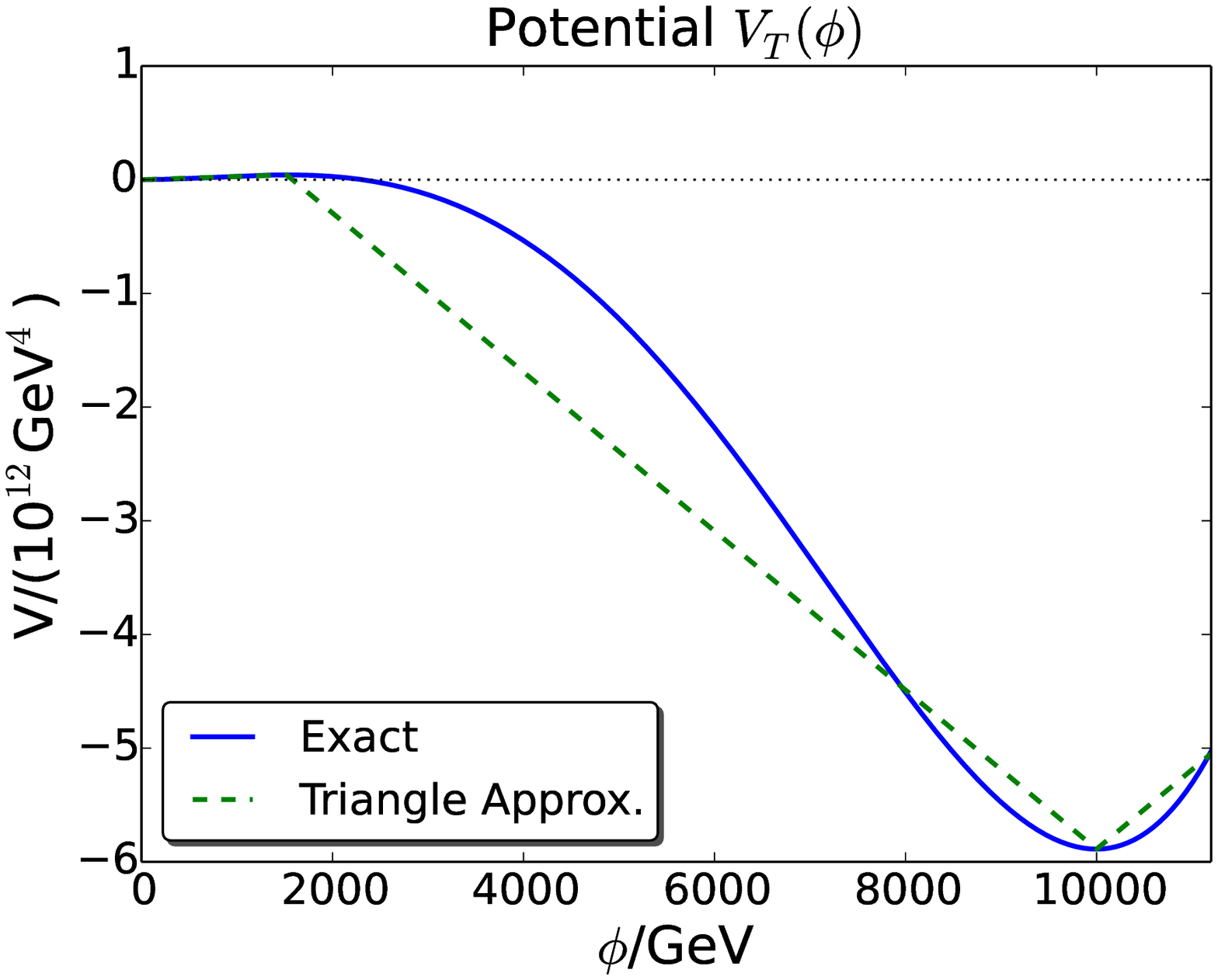}
\includegraphics[scale=0.38]{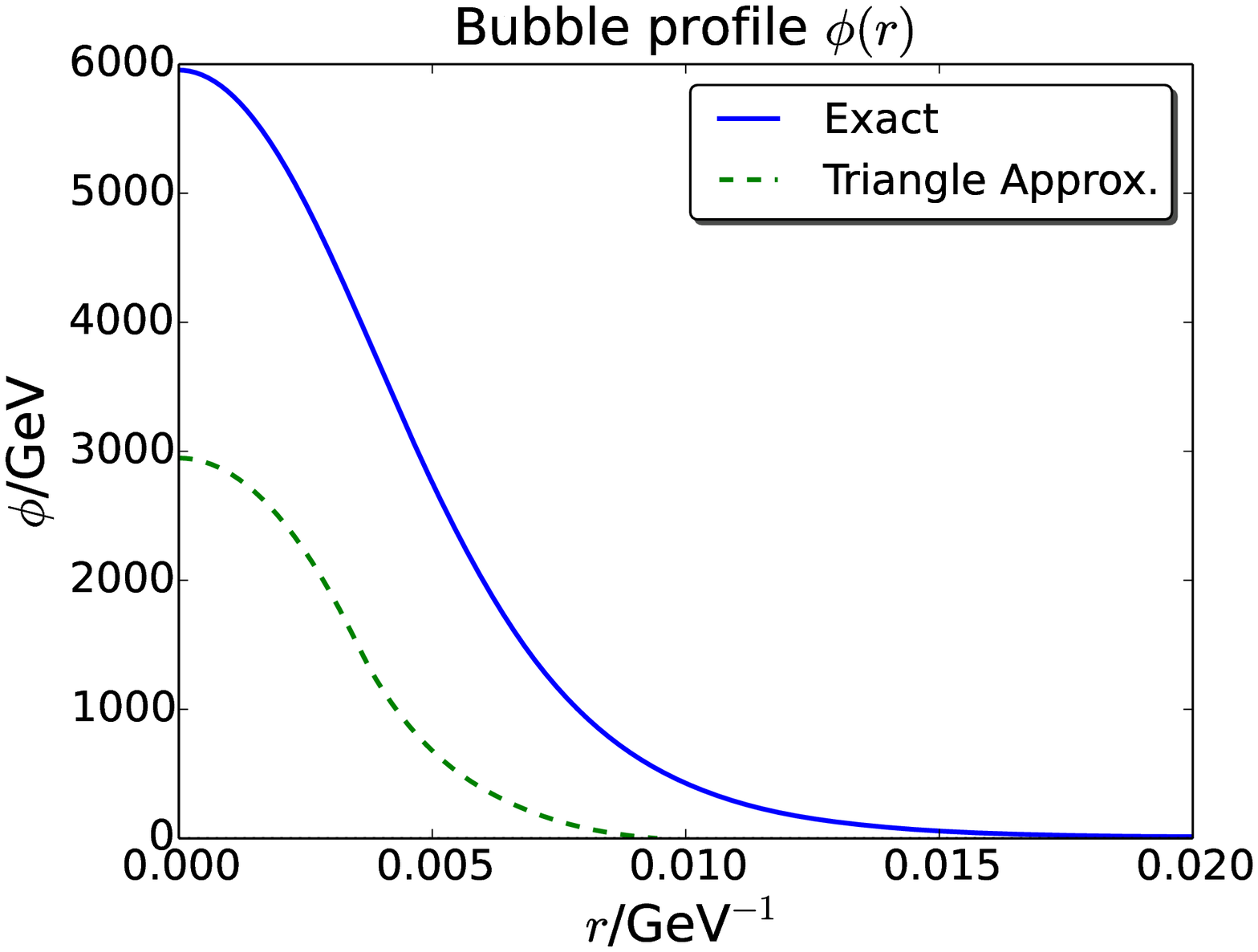}
\caption{Scalar potentials (left panel) and bubble profiles (right panel) of the model with $g_D = 0.6$ and $w = 10$~TeV at $T=800$~GeV by the exact numerical calculation (blue solid curve) and the triangle approximation (green dashed curve).}
\label{cmp2}
\end{figure}

The triangle approximation applied in Ref.~\cite{Jaeckel:2016jlh} can be illustrated in the upper panel of Fig.~\ref{cmp2} for the finite-temperature potential $V_T$ with $g_D = 0.6$, $w = 10$~TeV at a temperature $T = 800$~GeV. In this approximation, the field values and the potential energy densities of the maximum and the two minima are kept to be the same as the exact potential, but they are connected by straight lines. Therefore, the triangle potential is only characterized by the slope $\lambda_p$ ($\lambda_m$) of the line connecting the potential maximum and the false (true) vacuum, as well as the corresponding field distance $\Delta \phi_p$ ($\Delta \phi_m$). The advantage of the triangle potential is that the corresponding tunnelling solution and on-shell action can be solved analytically~\cite{Duncan:1992ai,Jaeckel:2016jlh}. It is particularly useful when the bubble profile is a thick wall. By defining the quantities:
\begin{eqnarray}
c=\frac{\lambda_m}{\lambda_p}\,, \quad a = (1+c)^{1/3}\,, \quad \kappa = \frac{\lambda_p}{(\Delta\phi_p)^3}\,,
\end{eqnarray}
the 3d action of such a thick-wall bubble can be written as follows\footnote{The on-shell thin-wall action in Eq.~(\ref{S3Tri}) is different from the one in Eq.~(35) of Ref.~\cite{Jaeckel:2016jlh}. But we regard that it is only a typo of Ref.~\cite{Jaeckel:2016jlh}, since the expressions and numerical results following Eq.~(35) seemed correct.}
\begin{eqnarray}\label{S3Tri}
S^T_3 = \frac{16\sqrt{6}\pi a^3 \Delta\phi_p }{5(a-1)^3 (1+2a)^{3/2} \sqrt{\kappa}}\,.
\end{eqnarray}
As pointed in Ref.~\cite{Jaeckel:2016jlh}, with the dark gauge coupling $g_D<1$, the nucleation bubble in this classically conformal model looks like a thick wall, as it could not reach the true vacuum at the bubble origin.

\begin{figure}[ht]
\includegraphics[scale = 0.38]{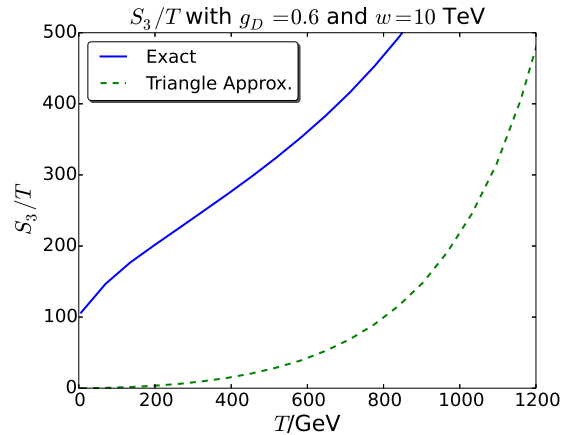}
\caption{$S_3/T$ as a function of temperature $T$ with the exact numerical calculation (blue solid curve) and the triangle approximation (green dashed curve). The parameters are chosen to be $g_X = 0.6$ and $w = 10$~TeV.}
\label{cmp}
\end{figure}

In the lower panel of Fig.~\ref{cmp2}, we show the tunnelling solutions to Eqs.~(\ref{DiffEq}-\ref{Bound}) by using the exact potential $V_T$ and its triangle approximation. It is obvious that the true bubble solution is much thicker and higher than its triangle-approximated counterpart. This feature can be traced back to the fact that the field value at the potential maximum is hierarchically smaller than that of the true vacuum, and the barrier around the top is broad but low-height. As shown in the upper panel, the triangle approximation underestimates the broadness near the potential top, which makes the tunnelling much easier. Therefore, the on-shell action under the triangle approximation is expected to be smaller than its exact value. This observation is especially true for a thick-wall bubble in the present model, since the bubble is only sensitive to the potential maximum region. This expectation is further supported by the calculations of $S_3/T$ as a function of temperature with both the exact numerical method and the triangle approximation in Eq.~(\ref{S3Tri}), as illustrated in Fig.~\ref{cmp}.
It is shown that exact numerical values of $S_3/T$ are always much larger than those with the triangle approximation, and the difference can be as large as 200. Consequently, it directly leads to distinct characteristics of the first-order PTs, with $T_n = 887$~GeV, $\alpha = 1.4\times 10^4$, and $\beta/H_* = 33.68$ for exact numerical calculations whereas $T_n \approx 60$~GeV, $\alpha = 0.26$ and $\beta/H_* = 172$ for the triangle approximation. With the triangle-approximated parameters, we can reproduce GW spectra in Figs.~6 and 7 of Ref.~\cite{Jaeckel:2016jlh}, which confirms that the inconsistency indeed comes from the application of the triangle approximation in Ref.~\cite{Jaeckel:2016jlh}. 

In summary, we have revisited the GW production during the first-order PT in a classically scale-invariant $SU(2)$ gauge sector studied by J.~Jaeckel, {\it et al.} in Ref.~\cite{Jaeckel:2016jlh}. Based on accurate numerical calculations of the nucleation bubble profiles and the 3d on-shell actions, we give in Fig.~\ref{GWCW} the GW spectra, which are shown different from those in Ref.~\cite{Jaeckel:2016jlh} in peak frequencies and spectrum shapes. We then argue that this inconsistency is mainly caused by the inappropriate use in Ref.~\cite{Jaeckel:2016jlh} of the triangle approximation, which greatly underestimates the broadness of the finite-temperature potential barriers when calculating thick-wall bubble actions. Note that this observation should be applied to more general nearly conformal dark sector~\cite{Konstandin:2011dr,Konstandin:2010cd}, since the finite-temperature potentials in this class of models have a common feature that the field value of the maximum is hierarchily smaller than that of the true vacuum and the barrier is broad but shallow. Moreover, we give in Eq.~(\ref{S3Tri}) the correct formulae for the 3d triangle-approximated on-shell action of a thick-wall bubble, rectifying a typo in Ref.~\cite{Jaeckel:2016jlh}. Finally, 
our precise calculations confirm the detectability of this model by the fifth phase of aLIGO, as pointed out by J.~Jaeckel, {\it et al.} However, our predicted sensitive GW frequencies correspond to those produced at $T_*\sim 30~{\rm PeV}$, rather than $T_* \sim 500$~TeV given in Ref.~\cite{Jaeckel:2016jlh}. 


We would like to thank Thomas Konstandin for useful discussions. DH is supported by the National Science Centre (Poland) research project, decision DEC-2014/15/B/ST2/00108, while BQL by the National Science Foundation of China under Grants No.~11335012 and No. 11475237.  


\end{document}